\documentclass[aps,prb,twocolumn,superscriptaddress,preprintnumbers,nofootinbib,showpacs]{revtex4}

\usepackage{graphicx}
\usepackage{amsbsy}
\usepackage{amsthm}
\usepackage{amsfonts}
\usepackage{amsmath}
\usepackage{amsopn}
\usepackage{amssymb}

\begin{document}


\title{Annealing Strategies in the Simulation of Fullerene Formation}



\author{Klaus Lichtenegger}
\email{klaus.lichtenegger@uni-graz.at}
\homepage{http://physik.uni-graz.at/~kll/}
\affiliation{Institute of Physics, Theory Division, Graz University, Austria}

\author{Wolfgang von der Linden}
\email{wvl@itp.tugraz.at}
\affiliation{Institute of Theoretical and Computational Physics,
Graz University of Technology, Austria}

\date{\today}

\begin{abstract}
  We investigate the formation of fullerene-like structures
  from hot Carbon gas using classical molecular dynamics,
  employing Brenner's potential. In particular we examine
  the influence of different annealing strategies on fullerene
  yield, which is characterized by the distribution of coordination
  numbers and polygon numbers. It will be shown that the fullerene yield strongly
  depends on the annealing strategy. Furthermore, we observe a close relation
  between polygon formation and the number of atoms surrounded by three atoms.
\end{abstract}


\maketitle

\section{Introduction\label{sec:intro}}

\subsection{Fullerenes and Nanotubes
\label{ssec:intro_fullerenes}}

Since their discovery~\cite{kroto1985} in 1985, fullerenes have been the subject
of intense studies, both experimentally and theoretically. These clusters have been
identified as roughly spherical carbon ``cages''. The fullerene family includes molecules
ranging from the (probably unstable) ${\rm C}_{20}$ to fullerenes which contain
hundreds or even thousands of carbon atoms. The most stable fullerenes are the
soccer ball-like ${\rm C}_{60}$ (also known as \emph{buckyball}) and the ${\rm C}_{70}$.

All fullerenes have in common the same principal structure: They are composed of
$sp^2$-hybridized carbon atoms, therefore being closely related to graphite. However,
since they are not planar, they can't be made up entirely of hexagons. The spherical
structure is typically achieved by insertion of pentagons. Also Carbon
nanotubes~\cite{Iijima91}, discovered 1991, have a similar structure -- the main
difference to fullerenes is the enormous aspect ratio (length vs. diameter).

Nanotubes can be interpreted as rolled-up graphite sheets, where the precise
procedure of rolling-up decides about the electronic structure of the nanotubes
(metallic or semiconducting). So in the absence of defects, nanotubes are composed
entirely of hexagons -- except the caps which close the tubes.

In the language of classical geometry, fullerenes can be described as polyhedrons,
so Euler theorem applies: For an arbitrary polyhedron with $V$ vertices, $E$ edges
and $F$ faces, the Euler characteristic $C=V-E+F$ is an invariant, for all simple
polyhedrons, $C=2$. From this theorem it can be deduced that any structure which
is topologically equivalent to a sphere and only composed of $n_6$ hexagons,
$n_5$ pentagons and $n_7$ heptagons must fulfill \vspace{-2mm}
\begin{equation}
  n_5-n_7 = 12
\end{equation}
with arbitrary $n_6$. So all fullerene molecules will contain exactly 12 pentagons
(unless there are heptagons present to compensate for a higher number of pentagons).
Therefore, $C_{20}$ is indeed the smallest possible fullerene, composed of 12
pentagons and no hexagons at all.

Since pentagons induce stress, structures which contain adjacent pentagons
show reduced stability (\emph{isolated pentagon rule}, IPR).
Obviously, for less then $60$ atoms there is no possibility to avoid this
occurrence, and also for $61\le n \le 69$ atoms no structure exists which
fulfills the IPR. In perfect icosahedral $C_{60}$ two pentagons are always
separated from each other by one hexagon, this is the most stable
configuration, followed by the rugby-ball like $C_{70}$ which again fulfills
the pentagon rule.

Further increasing the number of atoms does not enhance stability.
Multiple adjacent hexagons show a graphite-like structure that is more
severely disturbed by the pentagons the bigger the molecule gets.
So the yield of those fullerenes (for example at arc discharges)
is again much smaller than that of $C_{60}$ or $C_{70}$.

\subsection{Molecular Dynamics Studies
\label{ssec:intro_moldyn}}

A natural tool to study the dynamics of fullerene systems is
molecular dynamics (MD), where the equations of motion
for the nuclei are integrated numerically\cite{RapMD}. The most
important difference between various types of MD simulations
is the way to implement the interaction between the nuclei.

Depending on whether the electronic structure is simulated by
ab-initio methods\cite{MarxHutter00, Tuckerman02abinitio, CarParrinello},
tight-binding methods~\cite{TBMDlecture} or by more or less elaborate
force fields, the accessible system sizes and simulation times vary
dramatically.

While there is a wide variety of force-field approaches, certain
forms are particularly well-suited for the simulation of (hydro-)carbon
systems: Building on the approach of Abell and Tersoff\cite{Abell85,
Tersoff86, Tersoff88prl, Tersoff88prb, Tersoff89},
Donald W. Brenner has developed a potential\cite{BrennerPotential,
Brenner02} for hydrocarbons that is closely related to the potentials obtained
in the embedded atom methods (EAM). It contains empirical
parameters which were fitted to reproduce the properties of hydrocarbons
as well as those of carbon in diamond and graphite.

Its first use was the study of chemical vapor deposition, but found widespread
use in various simulations of carbon-based materials. This potential has
been incorporated in the BrennerMD code\cite{BrennerMD}, which was
used for the simulations of sec.~\ref{sec:simulation}.

MD studies of fullerenes by now encompass a broad range of topics,
among others
\begin{itemize}
\item studies of nanotube growth\cite{TBMDBoronAssist, GrowthEnergetics,
  NanoGrowth95, CatGrowth, XiGrowthDefect},
\item charge structure of nanotube growth from small clusters\cite{Melker},
\item nanotube cap formation and energetics~\cite{StabCapSWCNT},
\item reactive collision of graphite patches\cite{NTNhornGraphPatch} which
  lead to formation of the nuclei of nanotube, nanocage and other structures,
\item collisions between fullerenes\cite{DefectsCollision, C60C60Coll} and
  between fullerenes and other molecules\cite{C60CollH2},
\item the impact of fullerenes on graphite materials~\cite{C70GraphiteColl},
\item formation of diamond crystallites inside nested carbon fullerenes.\cite{DiamNucl}
\end{itemize}

A more comprehensive overview has been given elsewhere~\cite{DAKL04}.
Most interesting in the context of the present article are those studies which deal
with formation and fragmentation of fullerenes or fullerene-like structures.

Fullerene stability and fragmentation mechanism have been investigated
with a variety of computational methods~\cite{hussien-2008}, in particular
with tight-binding methods. There $\mathrm C_{60}$ turned out to be stable
against spontaneous disintegration for temperatures up to
$5000\;\mathrm{K}$\cite{Wang1992}. Similar results have been
obtained for other $\mathrm{C}_n$-systems.\cite{SystStudyStab, ThermalDisintegration, Zhang1993b}.
Most common fragmentation products are dimers, rings and chains; the
fragmentation temperature first increases linearly with cluster size, but becomes
nearly constant for fullerenes composed of 60 or more atoms.

Other simulations\cite{Kim1993} find structural changes in
$\mathrm{C}_{60}$ and C$_{70}$ between $3000$ and $4000\;\mathrm{K}$
and bond breaking around $5000\;\mathrm{K}$. For the melting
and evaporation of $C_{20}$, $C_{60}$ and $C_{240}$ one
finds\cite{MeltingFullerenes} a transition from the low-temperature
solid to a floppy ''Pretzel phase'', melting at about $T=4000$K and
conversion to carbon chain fragments when increasing the
temperature to $T\approx 10^4$K.

A notable feature of disintegration is the ejection of small clusters\cite{TBMDAnnealFrag}
(in particular $\mathrm{C}_2$, sometimes also $\mathrm{C}_3$)
already below the melting temperature. Similar results for Desintegration
and cage formation have been obtained using the Tersoff\cite{Marcos1999} or
Brenner\cite{Marayuma1998} potential.

When, studying fullerene formation and using monocyclic, bicyclic, and tricyclic
rings as precursors\cite{ClassMDFUllForm}, the most efficient temperature for the
formation of the cage-shaped structure from a ring is about 3000 K.
Also the growth process of higher fullerenes through adduction of small
carbon clusters and carbon atoms has been studied\cite{ContGrowth}: In this case
small clusters and single atoms easily adsorb on the surface of fullerene cages
which have defects, on collisions at thermal velocities; during an annealing process,
the attached clusters are soon incorporated into the network of the fullerene cages.

\bigskip

Summarizing these results, most studies find that during the annealing process,
one has a sequence of typical steps:

\begin{itemize}
\item For very high temperatures, one has a gas of single atoms or dimers.\\[-13pt]
\item When cooling down, the dimers tend to form chains that stay short first, but grow to length of
  $\mathcal{O}(10)$ atoms below $T=7000$K. This phase
  (``Pretzel phase''\cite{MeltingFullerenes}) can be interpreted as
  liquid-like.\footnote{Note that this phase is not
  always found\cite{hussien-2008}. We suspect that this is an artifact of
  the force field implementation, as for example a fixed neighbour list. Such
  a constraint makes the formation of long chains extremely unlikely
  since (in the case of 60 atoms) for a given dimer only four out of 58 atoms
  are capable of attaching to this dimer.}\\[-13pt]
\item Below $T \approx 5500$K most chains begin to form vaguely spherical
  structures where most atoms establish an additional bond to reach
  carbon's favored coordination number of three. In particular,
  hexagons, which are sign of locally graphite respectively
  fullerene-like structures, form.\\[-13pt]
\item The precise transition temperature for the
  change to fullerenes of course depends on the model employed for
  interaction, but the growth and removal of defects seem to be most
  efficient around $T=3000\;\mathrm{K}$.\\[-13pt]
\end{itemize}

The resulting structures are typically graphite- or fullerene-like with lots of defects.
The number of defects can be reduced by slow annealing, and it has been
possible (e.g. by isolating a fullerene precursor) to obtain perfect fullerene
molecules in some simulations.

All these simulations still have a severe drawback,
the times accessible by these simulations 
are still short compared to the timescales relevant for most physical processes
relevant for the  formation of fullerenes,
which is supposed to happen at $\mathcal{O}(s)$.  Simulations, however, can reach
in the case of rather small systems with
severely restricted types of force fields at most $\mathcal{O}(100\,\mathrm{ns})$.
In the present paper we address the question, whether
this shortcoming can be compensated by the
choice of an appropriate annealing strategy.

\subsection{Organization of the Paper
\label{ssec:intro_organization}}

The paper is organized as follows: After the introduction to fullerenes
(sec.~\ref{ssec:intro_fullerenes}) and corresponding molecular dynamics
simulations (sec.~\ref{ssec:intro_moldyn}), we turn to the basic simulation
setup, described in section~\ref{sec:simulation}.

Our analysis tools are described in section~\ref{sec:analysis}, where we discuss
distribution of coordination number  in sec.~\ref{ssec:analysis_coord} and polygon numbers
in sec.~\ref{ssec:analysis_polygon}.
In sec.~\ref{sec:Results} we present and discuss our results;
in sec.~\ref{sec:Summary} we summarize them and give a brief outlook.

\begin{figure*}[tbh]
\begin{center}
  \includegraphics[width=17cm]{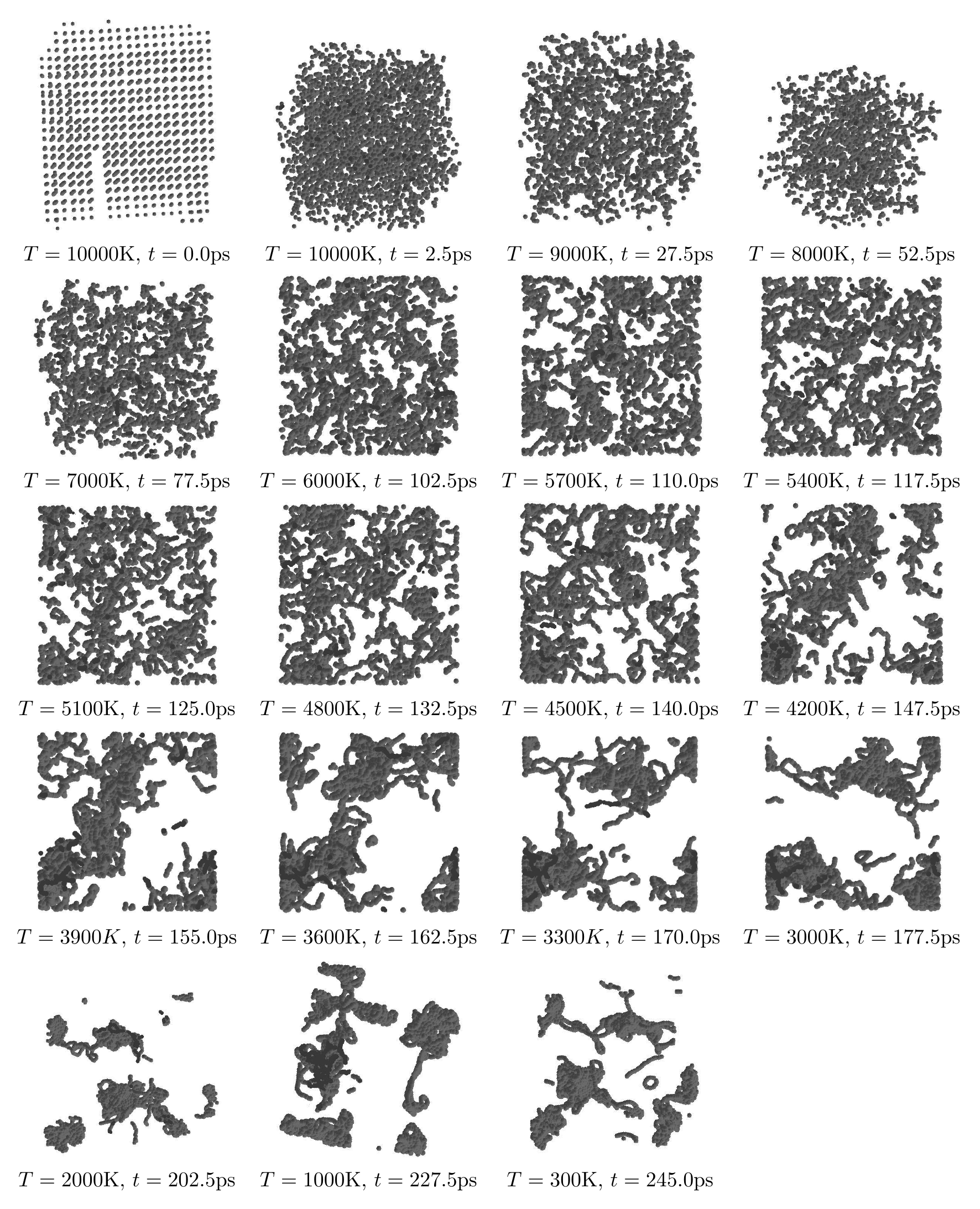}
\end{center}
\caption{Annealing hot carbon gas: snapshots. This simulation
  is similar to the ones performed with annealing strategy $T_1$,
  explained in sec.~\ref{sec:simulation} and illustrated
  in fig.~\ref{fig:annstrat}. One can clearly recognize both
  fullerene-like and chain structures.
  \label{fig1:annealing}}
\end{figure*}

\clearpage 

\section{Simulation Setup
\label{sec:simulation}}

Even with modern computing power, the times accessible to MD simulations
are still short compared to the timescales relevant for most physical processes
to be studied by this method. This also applies to the formation of fullerenes,
which is supposed to happen at $\mathcal{O}(\mathrm{s})$ while simulations are typically
limited to $\mathcal{O}(100\,\mathrm{ps})$, since one elementary MD step
corresponds to $\mathcal{O}(\mathrm{fs})$.
At best (studying small systems with
cheap and thus severely restricted types of force fields) one can reach
$\mathcal{O}(100\,\mathrm{ns})$.

One might expect that this shortcoming can be partially compensated by the
choice of an appropriate annealing strategy, e.g. if repeated heating-cooling
cycles can significantly improve the fullerene yield -- and this is the main question
to be studied in this article.

\subsection{Examples for Annealing Simulations
\label{sec:simulation_annealexamp}}

We have simulated the annealing of carbon gas from $T\approx 10^4$K to
room temperature. The simulations have been performed using the BrennerMD
code\cite{BrennerMD}, which makes use of Langevin dynamics, employing the
Fungimol graphical user interface\cite{Fungimol}.

An example for such type of simulation is shown in
fig.~\ref{fig1:annealing} (4024 carbon atoms in a box of
$100\,\mathrm{\AA} \times 100\,\mathrm{\AA} \times 100\,\mathrm{\AA}$
with a simulation time of $t=245$ps). The cooling was done in decrements
of $100\,\mathrm{K}$ with 5000 thermalization timesteps of $0.5\,\mathrm{fs}$
length in between. So the total simulation lasted
$t_{\mathrm{total}}=245\,\mathrm{ps}$ plus a few
femtoseconds of initialization steps without thermostat.

The typical stages discussed in sec.~\ref{ssec:intro_moldyn} are
clearly visible here: the dimer stage, the liquid-like chain phase,
the emergence of triple-bound atoms which tend to form planar
structures and thus the formation of fullerene-like structures.

Of course many defects occur even in the graphite- or
fullerene-like structures; a significant amount of atoms is
still contained in long chains even after cooling to room
temperature. These chains have (at least for even numbers
of atoms) higher ground-state energy than ring-based
structures (including graphite, graphene and fullerenes),
but they seem to form easier and are favorable at high
temperatures due to entropic reasons.

\begin{figure}
  \includegraphics[width=7.8cm]{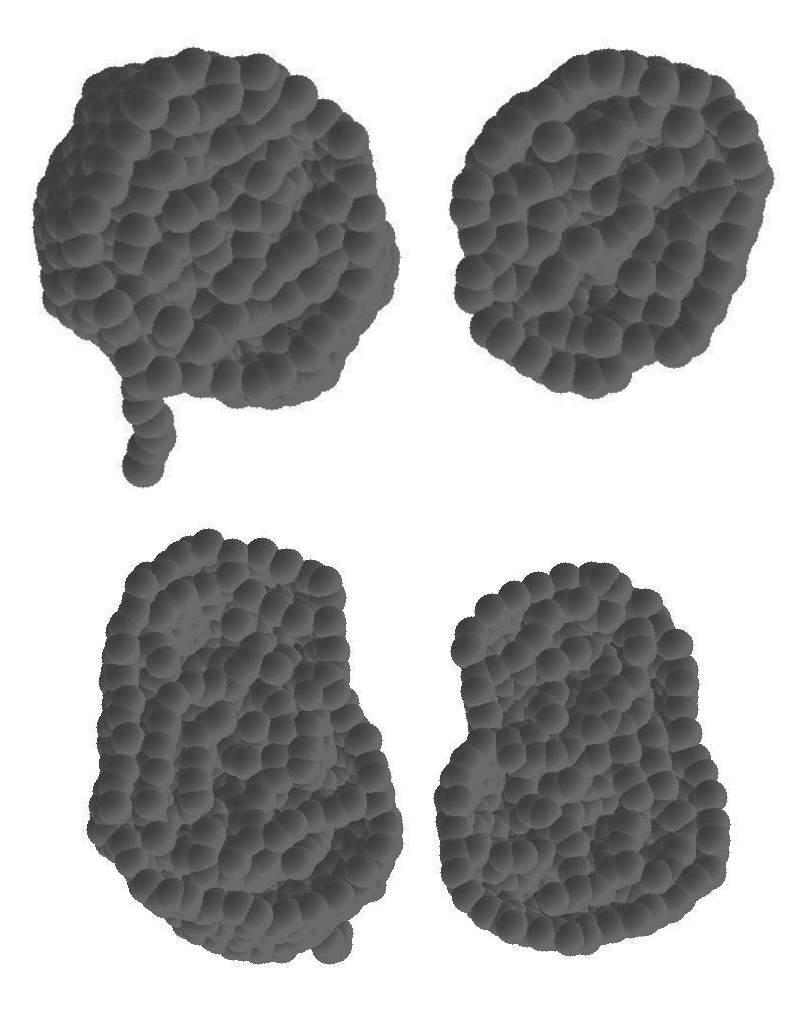}
  \caption{Cut through two final structures obtained by repeated heating
    and cooling (for two simulations starting with a gas of $N=1620$ atoms)
    \label{fig:simex_heatcool}}
\end{figure}

During the annealing process they should convert to smaller rings,
but here this is often impossible because of the rapid cooling process.
There are some similarities to the formation of an amorphous state
via rapid cooling. If -- in a very similar setup -- the monotonic decrease
of temperature is replaced by repeated heating an cooling cycles, the
final state typically looks like the one illustrated in
fig.~\ref{fig:simex_heatcool}.

One still has a considerable amount of long chains, but now they are
wound up and enclosed by roughly fullerene-like graphitic cages. The
resulting structures can be interpreted as partially frozen droplets with
a solid shell, but a liquid core.

\subsection{Setup for the Main Simulation
\label{sec:simulation_setupmain}}

In order to characterize the influence of annealing strategies in a
systematic way, we have repeatedly performed several simulations,
employing different annealing strategies. In all these simulations
the system consists 720 carbon atoms, initially distributed on
a simple cubic lattice in a box of
$80\,\mathrm{\AA} \times 80\,\mathrm{\AA} \times 80\,\mathrm{\AA}$.

\begin{figure}
  \begin{center}
    \includegraphics[width=8cm]{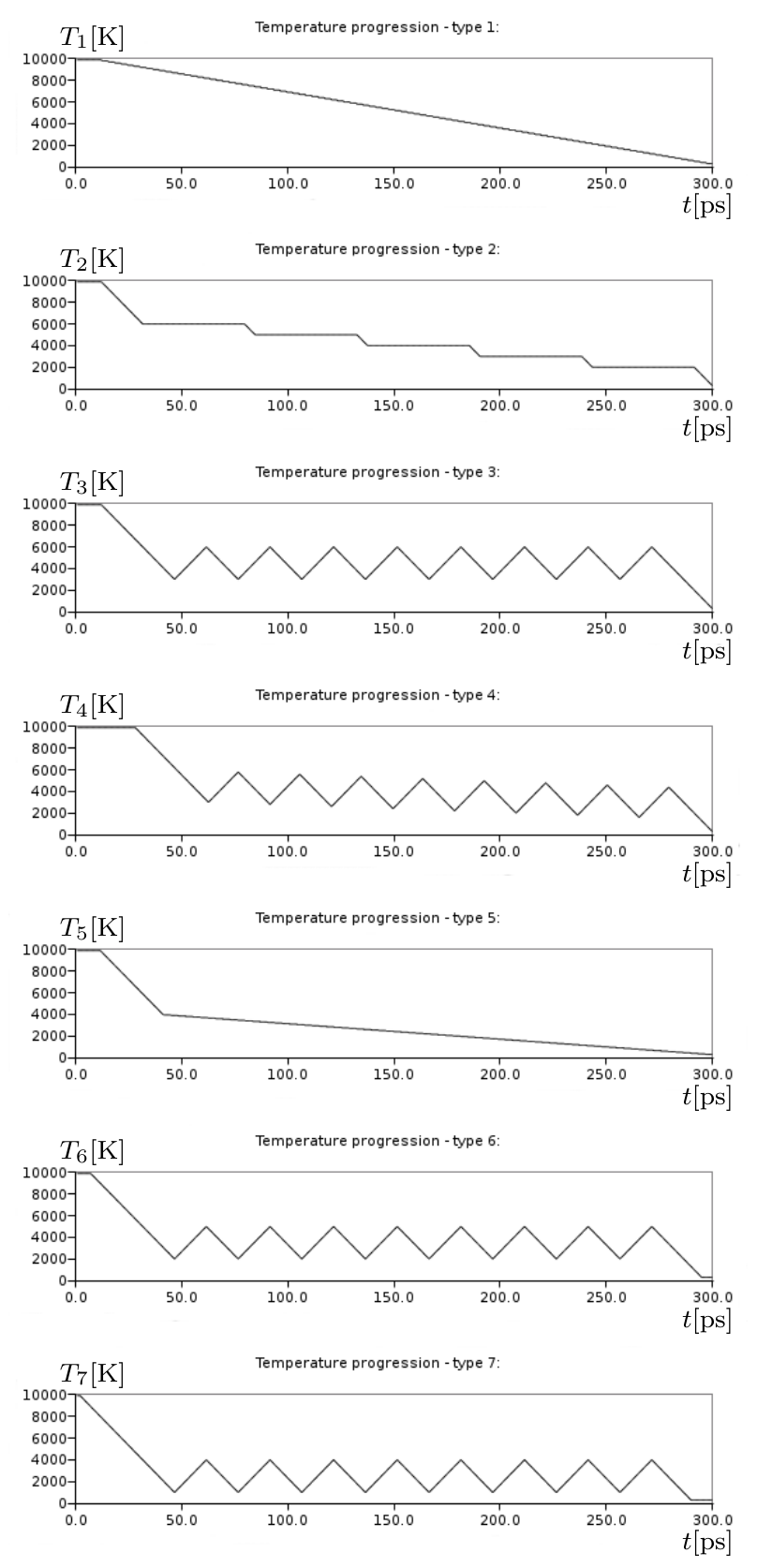}
  \end{center}
  \caption{Annealing strategies $T_1$ to $T_7$
    as described in sec.~\ref{sec:simulation_setupmain}.}
  \label{fig:annstrat}
\end{figure}

All simulations start with some initialization steps at
$T_{\rm initial}=9900\,\mathrm{K}$ and end at the same final temperature $T_\text{final}$. In all cases, the total simulation time was
$t=300\,\mathrm{ps}$ (600\,000 timesteps of $0.5\,\mathrm{fs}$).
Data were collected at the beginning of the simulation and every
$5\,\mathrm{ps}$ thereafter. Seven different annealing strategies
have been employed, which are also depicted in fig.~\ref{fig:annstrat}:

\begin{enumerate}
\item[($T_1$)] After an initialization period of $24\,000$ steps (corresponding to a
  simulation time of $12\,\mathrm{ps}$) at $T_\text{initial}$K, the temperature is
  lowered by $\Delta T=100\,\mathrm{K}$ every $6\,000$ timesteps
  (corresponding to $3\,\mathrm{ps}$) until $T_{\rm final}$
  is reached.
\item[($T_2$)] After the same initialization period as in $T2$, the system is rapidly
  cooled ($\Delta T=-100\,\mathrm{K}$ every $1\,000$ steps) down
  to $T=6000\,\mathrm{K}$ at which the system remains for $97\,000$
  steps. This process (rapid cooling by $1\,000\,\mathrm{K}$ in
  $5\,\mathrm{ps}$, followed $48\,\mathrm{ps}$ at constant temperature)
  is repeated until $T=2000\,\mathrm{K}$ is reached; afterwards
  rapid cooling is continued down to $T_\text{final}$.
\item[($T_3$)] After the same initialization period as in $T2$, the system is rapidly
  cooled ($\Delta T=-100\,\mathrm{K}$ every $1\,000$ steps) down
  to $T=3000\,\mathrm{K}$. After that, the temperature is rapidly
  ($\Delta T=\pm 100\,\mathrm{K}$ in $0.5\,\mathrm{ps}$) raised
  to $T=6000\,\mathrm{K}$ and lowered again down to
  $T=3000\,\mathrm{K}$. This heating-cooling cycle is repeated
  in total eight times; afterwards the system is further cooled down
  ($\Delta T=-100\,\mathrm{K}$ in $0.5\,\mathrm{ps}$) to
  $T_\text{final}$.
\item[($T_4$)] After an initialization period of $56\,000$ steps ($28\,\mathrm{ps}$),
  the system is rapidly ($\Delta T=-100\,\mathrm{K}$ in $0.5\,\mathrm{ps}$)
  cooled down to $T=3\,000\,\mathrm{K}$, heated to $T=5\,800\,\mathrm{K}$,
  cooled to $T=2\,800\,\mathrm{K}$, heated to $T=5\,600\,\mathrm{K}$,
  cooled to $T=2\,600\,\mathrm{K}$ and so on, until, after heating from
  $T=1\,600\,\mathrm{K}$ to $T=4\,400\,\mathrm{K}$ the system is
  cooled down to $T_\text{final}$.
\item[($T_5$)] After an initialization period of $23\,000$ steps, the system is
  rapidly cooled ($\Delta T=-100\,\mathrm{K}$ every $1\,000$ steps) down
  to $T=4000\,\mathrm{K}$. After that it is cooled slowly
  ($\Delta T=-50\,\mathrm{K}$ every $7\,000$ steps)
  down to $T_\text{final}$.
\end{enumerate}

\noindent The last two strategies are just variations of $(T3)$, but the heating-cooling
cycles operate between $T=2000\,\mathrm{K}$ and $T=5000\,\mathrm{K}$
for (T6) respectively between $T=1000\,\mathrm{K}$ and $T=4000\,\mathrm{K}$
for (T7). The annealing curves for theses seven strategies are displayed in
fig.~\ref{fig:annstrat}.

\section{Analysis Tools
\label{sec:analysis}}

In this article we focus on two characteristic quantities: the distribution of
coordination numbers and types of polygons present in the system.

\subsection{Coordination Numbers}
\label{ssec:analysis_coord}

The coordination number $z$ (number of direct neighbors) is
characteristic for the phase of carbon. A gas of carbon atoms has
$z=0$, for carbon dimers one has $z=1$. Carbon chains
and isolated rings have $z=2$. For the $sp^2$-hybridized carbon
atoms in graphite, graphene, nanotubes and fullerenes one has $z=3$,
while diamond structures are characterized by $z=4$.

The bond length of $sp^2$-hybridized carbon in graphite layers is $1.42\AA$,
for $sp^3$-hybridized atoms in diamond one has $1.54\AA$. The
carbon-carbon bond lengths in $C_{60}$ are $1.40\AA$ for adjacent hexagons
and $1.45\AA$ in pentagons. Similar values one has to expect in other fullerenes
or nanotubes. Therefore setting up a threshold of $a^*:=1.60\AA$ for the carbon-carbon
bond seems to be a reasonable choice. In our simulations, bonds are attached to all pairs of atoms that have a distance smaller than the threshold $a^*$. 
While other values for this threshold have some
influence on the results, this effect is small within reasonable limits.
We will use a normalized coordination number count
\begin{align*}
    q_{z} :=\frac{N_{z}}{N}\;,
\end{align*}
where $N_{z}$ is the number of atoms with coordination number $z$ and  $N$ is the total number of atoms.

\subsection{Number of Polygons
\label{ssec:analysis_polygon}}

Also the number of polygons present in the system is characteristic. Fullerenes
are mostly composed of hexagons and several pentagons, whereas pure carbon
nanotubes as well as graphite and carbon fibers are made from hexagons only.
Heptagons may break the convexity of fullerene structures, they repeatedly
appear as defects during cooling processes.

Simulations will be characterized by the normalized number of polygons
\begin{align*}
    q_\alpha &:=\frac{\text{number of polygons of type }\alpha}{N/2}\;.
\end{align*}
The polygon type can be pentagons ($\alpha=\mathrm{p}$), hexagons ($\alpha=\mathrm{h}$), or heptagons ($\alpha=7$).
The number of polygons is normalized to the graphene limit of hexagons, which is $\frac N2$ if there are  $N$ atoms.

\begin{figure}
  \begin{center}
    \includegraphics[width=8cm]{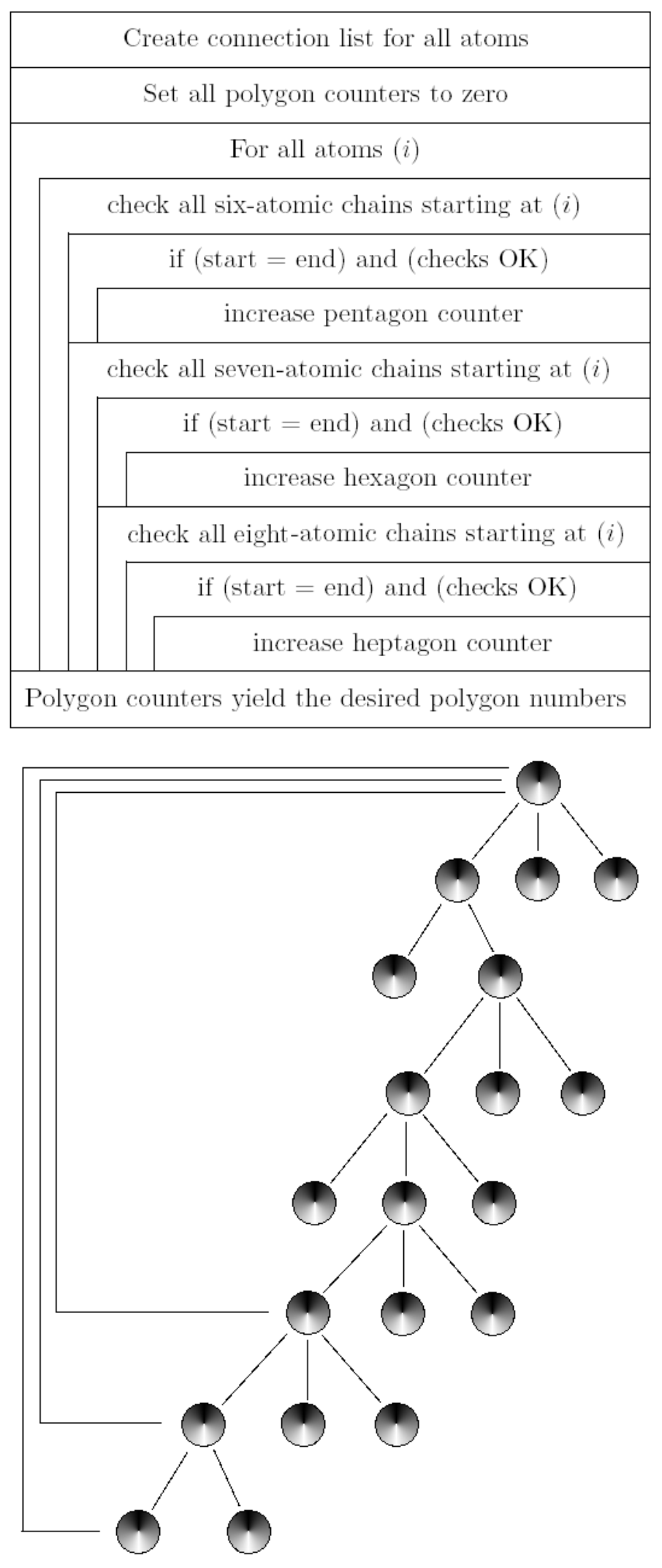}
   \end{center}
  \caption[Tree-based polygon search]{Tree-based polygon search: basic algorithm;
    graphical representation of six-, seven- and eight-atomic paths}
  \label{fig:TreeSearch}
\end{figure}

In order to determine the number polygons of length $L$, we use
a ring number algorithm which is built on a tree-search
as depicted in fig.~\ref{fig:TreeSearch}.

The idea is as follows. For each atom the algorithm checks  whether it is part of
a ring of length $L$ ($L=5,\,6,\,7$). The algorithm is repeated for the various
values of $L$ and for all atoms. Starting from the initial atom, the list of nearest
neighbor (n.n.) atoms is determined. The n.n. list contains those atoms which are
not further away from the reference atom than the threshold for the bond length.

In turn, for each atom in the n.n. list the corresponding n.n. list is generated.
The procedure is repeated $L$ times. The resulting structure can be regarded
as a graph with atoms as nodes and bonds as edges.

So, roughly speaking, for each particle all possible paths of length $L$ starting
from the selected atom and leading in each step to one of the
nearest neighbors of the current atom, are determined. From all these paths,
those for which starting and end point coincide (closed paths) are selected.

\begin{figure*}
  \begin{center}
    \includegraphics[width=17cm]{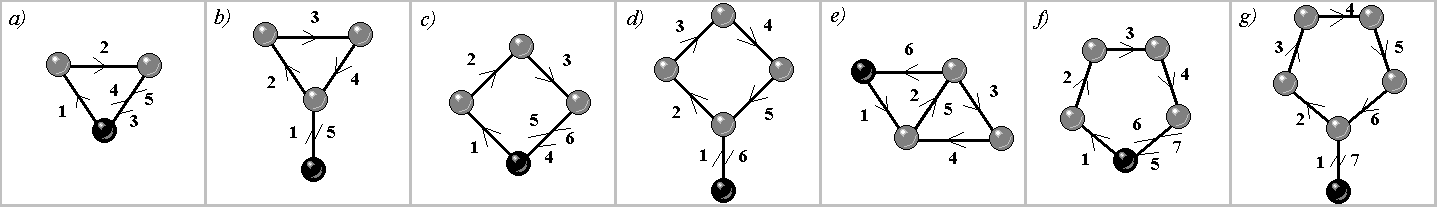}
  \end{center}
  \caption[Unwanted polygon findings]{Unwanted polygon findings:
    (a), (b) false pentagons; (c), (d), (e) false hexagons; (f), (g) false heptagons}
  \label{fig:FalsePolygons}
\end{figure*}

Next we eliminating those structures, in which bonds are repeatedly traversed
(as shown in fig.~\ref{fig:FalsePolygons}a and \ref{fig:FalsePolygons}b).
When all particles have been checked, the resulting number of detected closed
paths of length $L$  equals $2L$ times the number of polygons in the system,
since each atom of a polygon is used as initial atom and for each atom there are
two possible paths, one clockwise, the other one counter-clockwise.

\section{Results and Discussion
\label{sec:Results}}

The goal of this paper is to reveal the details of the polygon formation in order to develop an improved annealing strategy.
To this end, MD simulations for the  different annealing strategies presented above are performed and analyzed.
In all simulations, the total number of atoms, initial and final temperatures, as well as the total number of MD steps is the same.

\begin{figure}
\begin{center}
  \includegraphics[width=8.5cm]{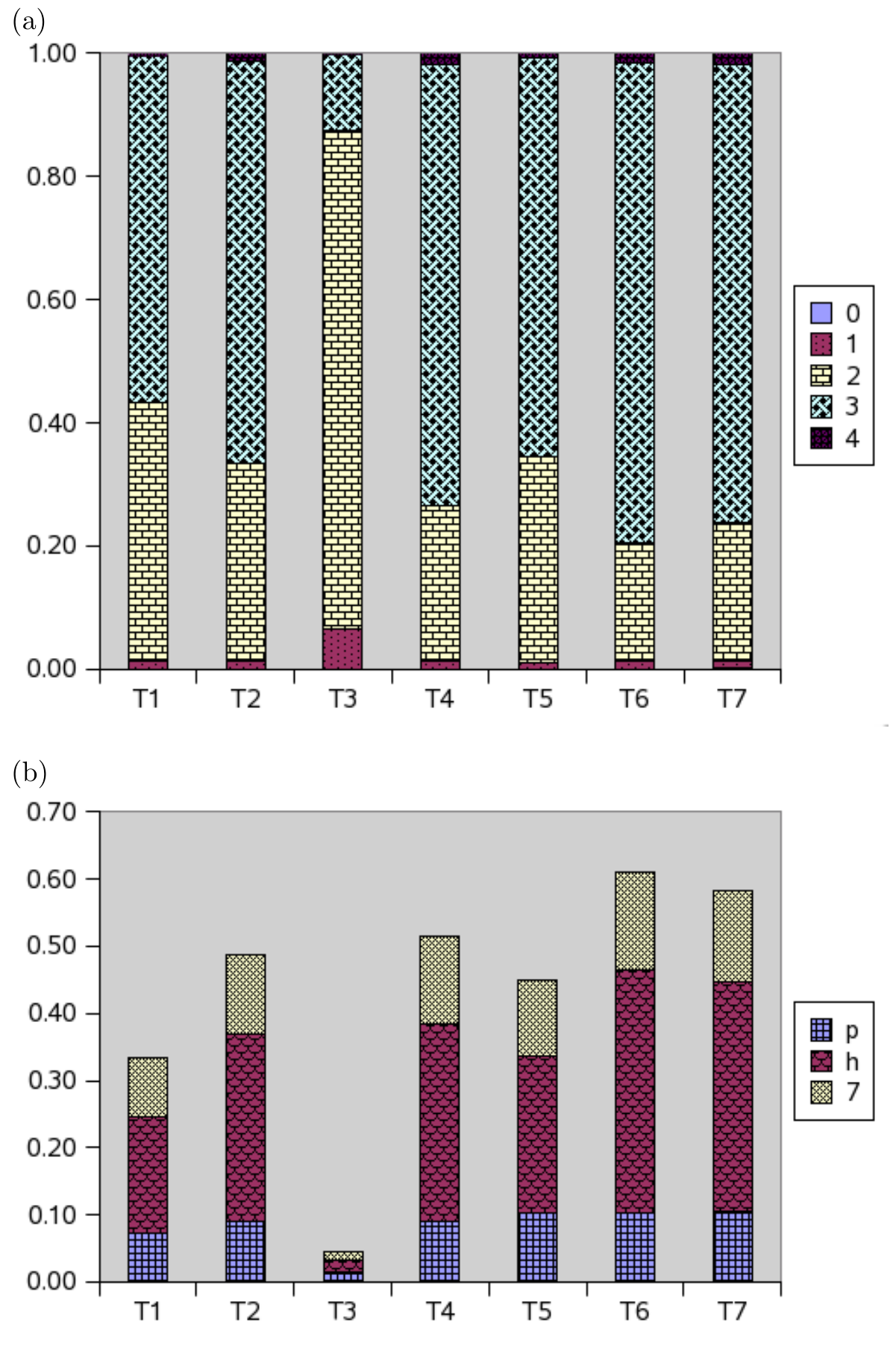}
\end{center}
\caption{Results: (a) normalized coordination number count $N_z$ for 
  $z=(0, 1, 2, 3, 4)$ after annealing runs $T_1$ to $T_7$;
  (b) number of polygons
  (pentagons p, hexagons h, heptagons 7),
  normalized to the graphene limit,
  for annealing runs $T_1$ to $T_7$}
\label{klfig:results}
\end{figure}

In figs.~\ref{klfig:results}a and~\ref{klfig:results}b the final results are summarized in form of the normalized coordination number count
 and the normalized number of polygons for the various annealing strategies.
An interesting observation of fig.~\ref{klfig:results}b is the fact that the ratio of the number of hexagons to pentagons and heptagons is almost the same in all cases. An exception is strategy $T_3$, in which the polygon yield is negligibly small in any case.

A comparison of fig.~\ref{klfig:results}a and~\ref{klfig:results}b reveals that the polygon yield is roughly proportional to the number of atoms with coordination number 3. This implies that polygons are not formed as isolated rings, but rather
immediately as locally connected planar structures. The ranking of the annealing strategies in view of the polygon yield is $T_3$, $T_1$, $T_2$, $T_5$, $T_4$, $T_7$, $T_6$. The most ineffective  strategy of all is clearly $T_3$ with almost no polygons, but long chains instead.

The expression
\begin{equation}
  N_{\rm full} = \frac{\text{number pentagons} - \text{number heptagons}}{12}
  \label{eq:numberfull}
\end{equation}
could be expected to characterize the number of fullerenes present
in a system, provided all carbon atoms are precisely bounded to
three other atoms.

In our simulations, however, we find the number of heptagons to be typically
larger than the number of pentagons. At second thought, this should not really be
surprising since in the ``chain phase'' the formation of longer rings is more likely than the formation
of shorter ones. It shows, however, that we are still quite far from a system which
only contains fullerenes.

\begin{figure}
  \begin{center}
    \includegraphics[width=8.5cm]{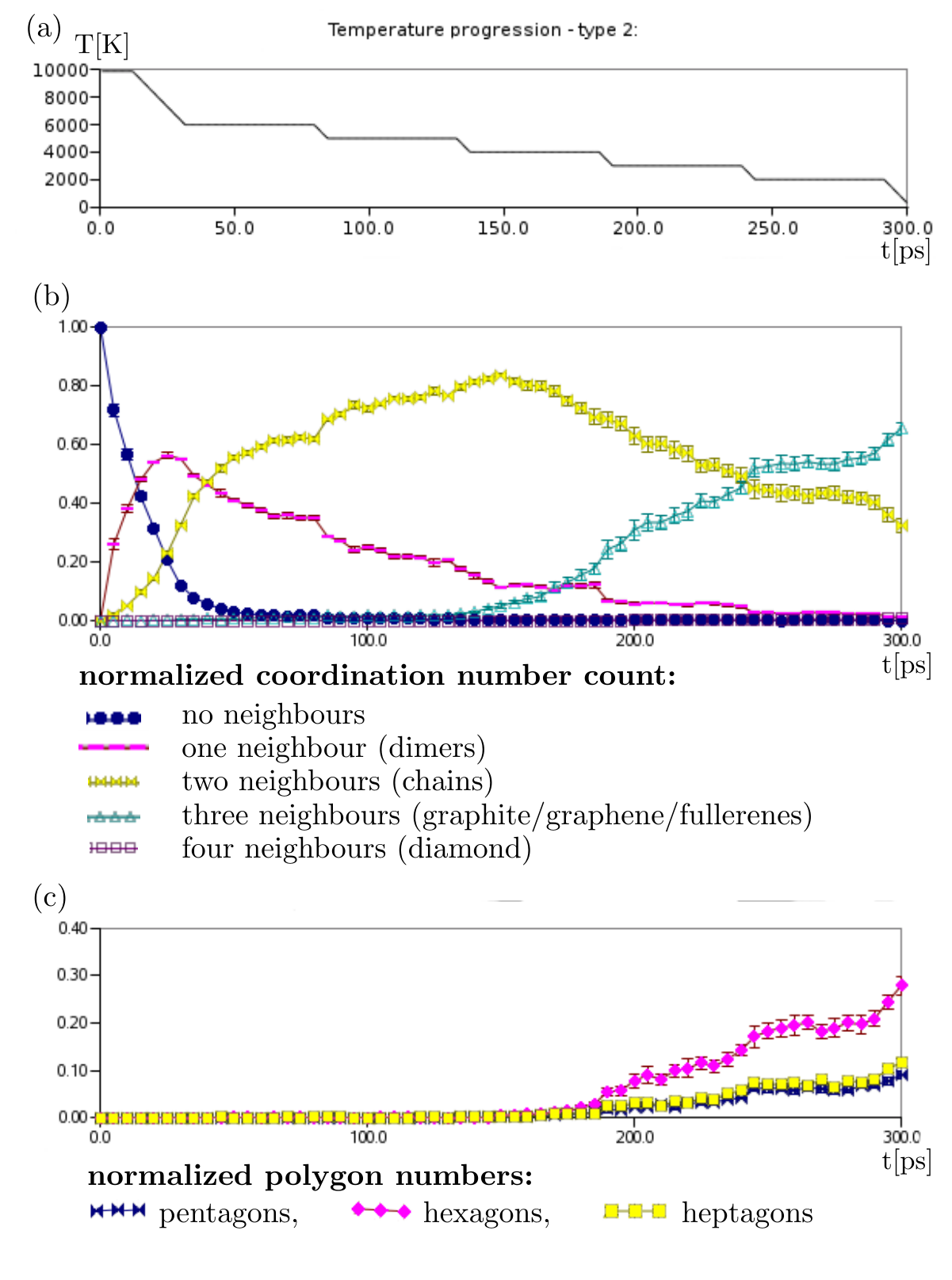}
  \end{center}
  \caption{Annealing strategy $T_2$: (a) temperature
  progression; (b) normalized coordination number count;
  (c) normalized number of polygons.}
\label{klfig:resultsT2}
\end{figure}

\begin{figure}
  \begin{center}
    \includegraphics[width=8.5cm]{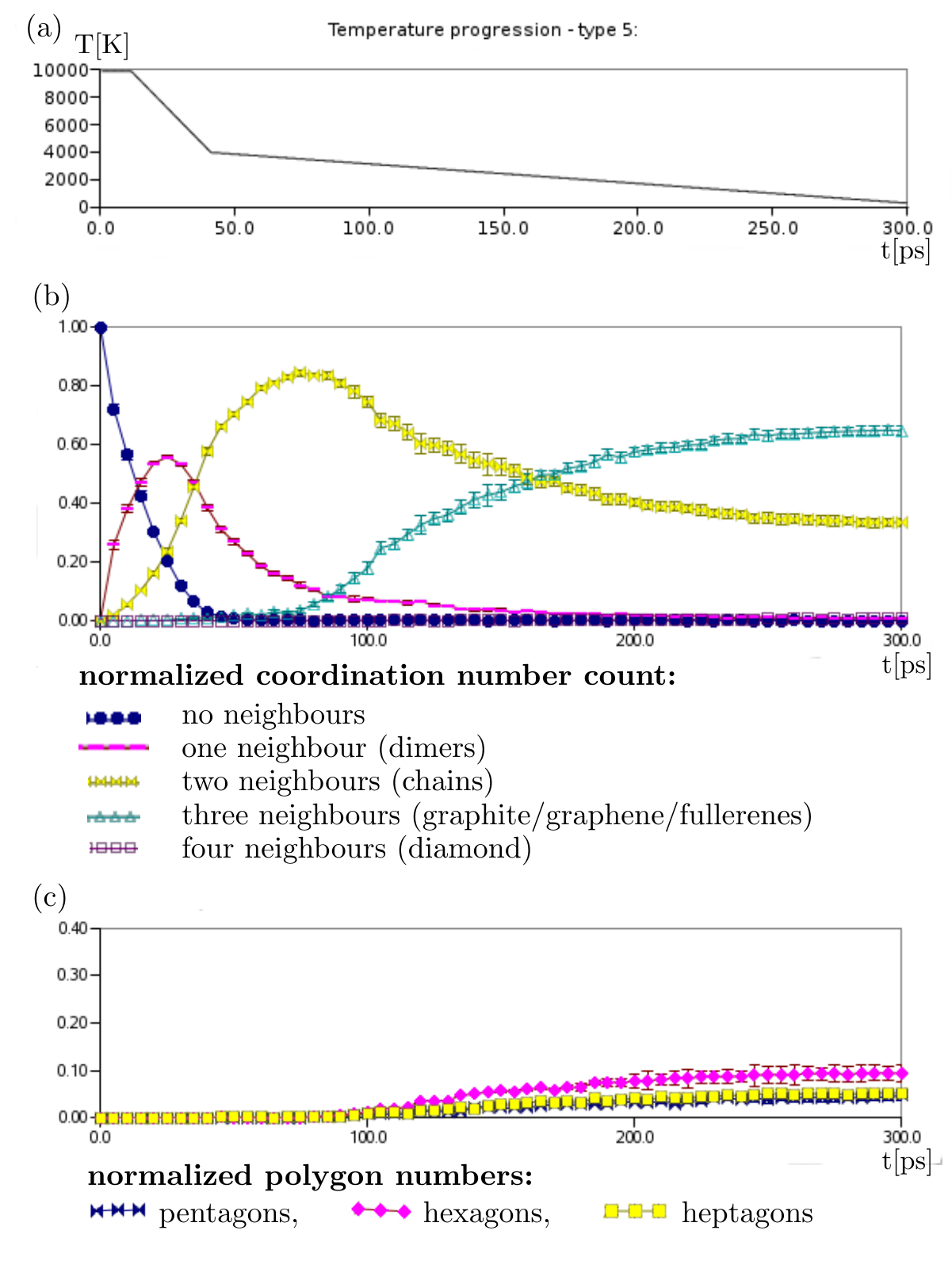}
  \end{center}
  \caption{Annealing strategy $T_5$: (a) temperature
  progression; (b) normalized coordination number count, 
  (c) normalized number of polygons.}
\label{klfig:resultsT5}
\end{figure}

\begin{figure}
  \begin{center}
    \includegraphics[width=8.5cm]{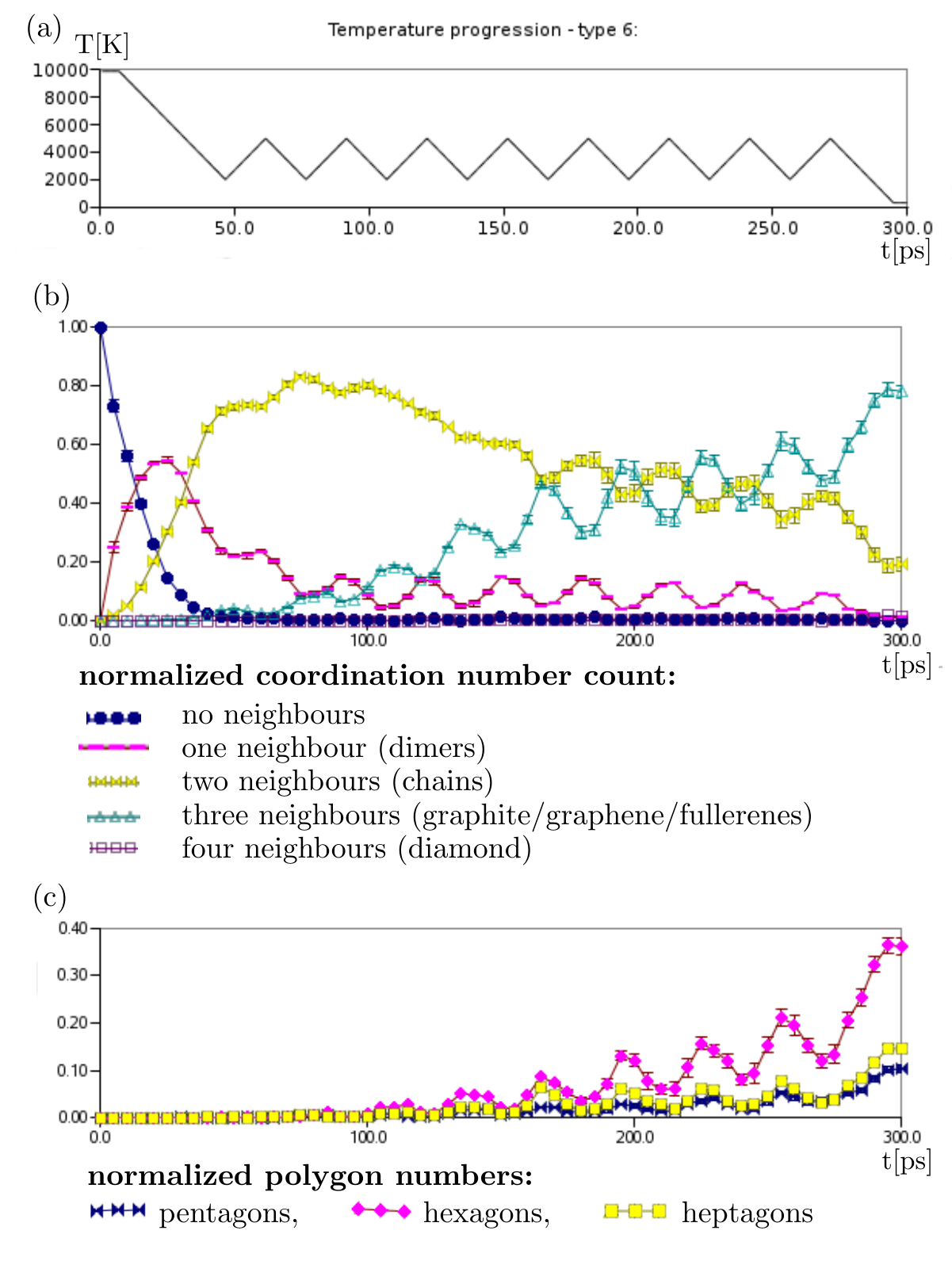}
  \end{center}
  \caption{Annealing strategy $T_6$: (a) temperature
  progression; (b) normalized coordination number count,
  (c) normalized number of polygons.}
\label{klfig:resultsT6}
\end{figure}

According to  fig.~\ref{klfig:results} the strategy $T_3$ has a very poor polygon yield, while the opposite is true for the best strategy $T_6$, although the two annealing strategies look fairly similar. The only but crucial difference is that in the first case the heating-cooling cycles operate between 3000~K and 6000~K, while in second case the temperatures are 2000~K and 5000~K. There are two effects which are expected to be responsible for that:

\begin{itemize}
\item In the temperature range of $(2000~\mathrm{K},\,3000~\mathrm{K})$ lies the crossover temperature $T^*_3$, at which the number of atoms with coordination number 3 start to dominate over those with coordination number 2. In this range the formation of graphite- or fullerene-like structures seems to be most efficient.
\item At the same time the heating-cooling cycles of $T_6$ do not exceed (contrary to the $T_3$) the onset temperature $T_o$ of polygon formation. As we see in fig.~\ref{klfig:resultsT6}, the system reacts almost instantaneously on temperature changes, therefore, heating above $T_o$ would break up the elementary building blocks of the polygons.
\end{itemize}

Fig.~\ref{klfig:resultsT2} contains the the detailed time evolution of the normalized coordination count and the normalized number of polygons during the anealing process for strategy $T_2$, in which a stepwise cooling is performed with  five intermediate temperatures, such that the system is almost in thermodynamic equilibration at the intermediate temperatures.
Based on the results in fig~\ref{klfig:resultsT2}a we define regions in which atoms of a specific coordination number dominate. We conclude that atoms with coordination number 2 dominate already at $T=10000$~K down to $T=6000$~K. From $T=6000$~K down to a crossover temperature $T^*_3=2000$~K, atoms with coordination number 2 predominate, while below $T^*_3$ atoms of coordinations number 3 are in the majority. In the observed temperature range, we found almost no atoms with coordination number 4, the prerequisite for the diamond structure.

The onset of polygon formation, according to fig~\ref{klfig:resultsT2}b, takes place at about $t_p=90$~ps, while the formation of atoms with coordination number 3 is roughly at $t_o=125$~ps. The polygon formation happens right after the cooling from $T=6000$~K to $T=5000$~K, while the formation of atoms with coordination number 3 takes place later at the same temperature. Hence, the polygons are formed first as ringlike structures before they combine locally to form planar structures. Interestingly, the opposite is the case if the system is cooled more rapidly without reaching thermodynamic equilibrium, as is observed in fig.~\ref{klfig:resultsT5}.

The onset of polygon formation in fig~\ref{klfig:resultsT5} is at a significant lower temperature than in
fig.~\ref{klfig:resultsT2}, due to the inertia of the system. Interestingly, the simulation time before polygon formation sets in is the same in both cases. The comparison of coordination number count  in both strategies reveals that the system passes through comparable phases, i.e. the depletion of isolated atom, the formation of dimers,  the increasing and decreasing number of atoms in chains, followed by the gradual formation of atoms with coordination number 3.


The best of all scrutinized strategies is $T_6$, where the systems passes through several heating-cooling cycles. The results are depicted in fig.~\ref{klfig:resultsT6}. Although the fraction of atoms with coordination number 3 seems converged in fig.~\ref{klfig:resultsT2}a for $T=2000K$ at a value of $q_{z=3}\approx 0.5$, the annealing in $T_6$ leads to a significant higher yield of about 0.7 when $T=2000K$ is reached during the last cooling cycle. At the same time the number of atoms with coordination number 2 is reduced by the same amount. Similarly, the number of polygons is roughly increased by $40\%$ in strategy $T_6$, when $T=2000$~K is reached during the last cooling cycle.

\begin{table}
  \begin{center}
  \begin{tabular}{|c||r|r|r|} \hline
    & 8 cycles & 16 cycles & 24 cycles \\ \hline \hline
    $q_{0}$ & 0\hspace{3mm} & 0\hspace{3mm} & 0\hspace{3mm} \\
    $q_{1}$ & 0.013\hspace{3mm} & 0.008\hspace{3mm} & 0.010\hspace{3mm} \\
    $q_2$ & 0.243\hspace{3mm} & 0.178\hspace{3mm} & 0.171\hspace{3mm} \\
    $q_3$ & 0.731\hspace{3mm} & 0.789\hspace{3mm} & 0.796\hspace{3mm} \\
    $q_4$ & 0.013\hspace{3mm} & 0.025\hspace{3mm} & 0.024\hspace{3mm} \\ \hline \hline
    $q_\text{p}$ & 0.119\hspace{3mm} & 0.083\hspace{3mm} & 0.078\hspace{3mm} \\
    $q_\text{h}$ & 0.358\hspace{3mm} & 0.444\hspace{3mm} & 0.442\hspace{3mm} \\
    $q_\text{7}$ & 0.152\hspace{3mm} & 0.122\hspace{3mm} & 0.128\hspace{3mm} \\ \hline
  \end{tabular}
  \end{center}
\caption{Effects of longer simulation runs
  on the fullerene yield for strategy $T_6$.
  The simulations have been performed for a specific intial configuration
  and eight, 16, 24 instead of eight)}
\label{tab:longruns}
\end{table}

Exemplary simulations with a strategy similar to $T_6$, but a larger number of heating-cooling cycles
yield the results given in table~\ref{tab:longruns}. The extension from eight to 16 cycles significantly increases the number of triple-bonded atoms (mostly at the expense of double-bonded ones) and the number of hexagons. A further extension to 24 cycles only marginally changes the results, so for this particular strategy we seem to have reached a plateau.

\bigskip

\section{Summary and Outlook
\label{sec:Summary}}

We have simulated the structures resulting from annealing a hot carbon gas
with different strategies, i.e. different forms of temperature evolution. The structures
have been analyzed with respect to the coordination number count 
and the number of pentagons, hexagons and heptagons.

We have found the yield of fullerene-like structures to be quite sensitive to the
strategy employed. Thus the fullerene yield can either be significantly enhanced
or almost reduced to zero by appropriate annealing strategies.

The best strategy included several heating-cooling cycles, operating below the
fullerene disintegration temperature (corresponding to the onset temperature of polygon
formation), but encompassing the crossover temperature between double- and
triple-bonded atoms

\smallskip

Since timescales in MD simulations are short compared to those of most physical
processes, an optimized annealing strategy can be an important means to obtain
more realistic results for MD simulations of fullerene formation. Alternatively,
repeated heating-cooling cycles which, performed in the optimal temperature window,
gave the  best fullerene yield, might also occur (on larger timescales) in nature
and enhance the fullerene yield e.g. in soot.


While some scrutinized strategies were obviously superior to uniform annealing,
it is not at all clear whether we have found an even close-to-optimal strategy in
order to maximize the fullerene yield. Here a stochastic optimization (employing
simulated annealing or genetic algorithms) of the annealing strategy might provide
valuable information on how far the limits can be pushed for short-time simulations.


\begin{thebibliography}{99}

\bibitem{kroto1985} H.~W. Kroto et~al., Nature \textbf{318}, 162, 1985

\bibitem{Iijima91} S. Iijima, Nature \textbf{354}, 56 - 58, 1991

\bibitem{FormProp}
   T. Oku, M. Kuno, H. Kitahara, I. Narita,
   Int. J. of Inorg. Mat. \textbf{3}, 597-612, 2001

\bibitem{DAKL04}
  K. Lichtenegger, \textit{Application of Molecular Dynamics Simulation Methods
  in the Nanoscale Regime}, master's thesis, Graz University of Technology, 2004;
  available on \texttt{http://physik.uni-graz.at/$\sim$kll/nanoMD.pdf}

\bibitem{RapMD}
  D. C. Rapaport, \textit{The Art of Molecular Dynamics Simulation},
  Cambridge University Press; Revised Edition, 2004

\bibitem{Ercolessi}
  F. Ercolessi, \textit{A molecular dynamics primer}, Trieste,
  \texttt{http://www.fisica.uniud.it/$\sim$ercolessi/md/}

\bibitem{MarxHutter00}
  Dominik Marx and J\"urg Hutter,
  NIC Series, Vol. 1, pp. 301-449, 2000, \texttt{http://www.fz-juelich.de/nic-series/}

\bibitem{Tuckerman02abinitio}
  Mark E. Tuckerman,
  J. Phys.: Condens. Matter 14 (2002) R1297-R1355

\bibitem{CarParrinello}
  R. Car,  M. Parrinello,
 Phys. Rev. Lett., Vol. 55 (1985) 22

\bibitem{TBMDlecture}
   see for example F. Ercolessi,
   \emph{Lecture notes on Tight-Binding Molecular Dynamics,
   and Tight-Binding justification of classical potentials},
   available on \texttt{www.fisica.uniud.it/$\sim$ercolessi/SA/tb.pdf}

\bibitem{Abell85}
  G. C. Abell, Phys. Rev. B 31, 6184 (1985).

\bibitem{Tersoff86}
  J. Tersoff, Phys. Rev. Lett. 56, 632 (1986).

\bibitem{Tersoff88prl}
  J. Tersoff, Phys. Rev. Lett. 61, 2879 (1988).

\bibitem{Tersoff88prb}
  J. Tersoff, Phys. Rev. B 37, 6991 (1988).

\bibitem{Tersoff89}
  J. Tersoff, Phys. Rev. B 39, 5566 (1989).

\bibitem{BrennerPotential}
  D. W. Brenner,
  Physical Review B \textbf{42}, 15, 1990

\bibitem{Brenner02}
  D. W. Brenner, O. A. Shenderova, J. A. Harrison, S. J. Stuart, and S. B. Sinnott,
  J. Phys.: Condens. Matter 14, 783 (2002).

\bibitem{BrennerMD}
  D. Brenner et al., \texttt{brennermd} code, available at
  \texttt{http://sourceforge.net/projects/brennermd/}

\bibitem{TBMDBoronAssist}
  E. Hern\'andez, P. Ordej\'on, I. Boustani, A. Rubio and J.A. Alonso,
  \texttt{arXiv:cond-mat/0006230}

\bibitem{GrowthEnergetics}
  A. Maiti, C.J. Brabec, C.M. Roland, J. Bernholc,
  Phys. Rev. Lett. 73 (1994) 18

\bibitem{NanoGrowth95}
  A. Maiti, C.J. Brabec, C. Roland, J. Bernholc,
  Phys. Rev. B 52 (1995) 20

\bibitem{CatGrowth}
  A. Maiti, C. J. Brabec and J. Bernholc,
  Phys. Rev. B 55 (1997) 10

\bibitem{XiGrowthDefect}
  Y. Xia, Y. Ma, Y. Xing, Y. Mu, C. Tan, L. Mei,
  Phys. Rev. B 15, 61 (2000) 16

\bibitem{Melker}
  A.I. Melker, S.N. Romanov, D.A. Kornilov,
  Mater.Phys.Mech.2 (2000) 42-50

\bibitem{StabCapSWCNT}
  D.-H. Oh, Young Hee Lee,
  Phys. Rev. B 58 (1998) 11

\bibitem{NTNhornGraphPatch}
  T. Kawai, Y. Miyamoto, O. Sugino, Y. Koga,
  Phys. Rev. B 66, 033404 (2002)

\bibitem{DefectsCollision}
  Y. Xia, Y. Mu, Y. Xing, C. Tan, L. Mei,
  Phys. Rev. B 56 (1997) 8

\bibitem{C60C60Coll}
  Y. Xia, C. Tan, Y. Mu, Y. Xing,
  Nucl. Instr. Meth. in Phys. Res. B 135 (1998) 195-200

\bibitem{C60CollH2}
  Y. Xia, Y. Xing, C. Tan, L. Mei,
  Phys. Rev. B 52 (1995) 1

\bibitem{C70GraphiteColl}
  Z. Man, Z. Pan, J. Xie, Y. Ho,
  Nucl. Instr. Meth. Phys. Res. B 135 (1998) 342-345

\bibitem{DiamNucl}
  R. Astala, M. Kaukonen, R. M. Nieminen, G. Jungnickel, T. Frauenheim,
  Phys. Rev.B 63, 081402 (2001)

\bibitem{hussien-2008}
  A. Hussien, A. Yakubovich, A. Solov'yov, W. Greiner,
  \texttt{arXiv.org:0807.4435}, 2008

\bibitem{Wang1992}
  C. Z. Wang, C. H. Xu, C. T. Chan, and K. M. Ho, J. Phys. Chem. 96, 3563 (1992).

\bibitem{SystStudyStab}
  B.L. Zhang, C.H. Xu, C.Z. Wang, C.T. Chan and K.M. Ho,
  Phys. Rev. B 46 (1992) 11

\bibitem{ThermalDisintegration}
  B. L. Zhang, C. Z. Wang, C. T. Chan, and K. M. Ho,
  Phys. Rev. B 48, 11381 (1993).

\bibitem{Zhang1993b}
  B. L. Zhang, C. Z. Wang, K. M. Ho, and C. T. Chan,
  Z. Phys. D 26, 285 (1993).

\bibitem{Kim1993}
  E. Kim, Y. H. Lee, and J. Y. Lee,
  Phys. Rev. B 48, 18230 (1993).

\bibitem{MeltingFullerenes}
  S. G. Kim, D. Tom\'anek,
  Phys. Rev. Lett. 72, 2418 (1994).

\bibitem{TBMDAnnealFrag}
  C. Xu, G. E. Scuseria,
  Phys. Rev. Lett. 72, 5 (1994) 31

\bibitem{Marcos1999}
  P. A. Marcos, J. A. Alonso, A. Rubio, and M. J. L\'opez,
  Eur. Phys. J. D 6, 221 (1999).

\bibitem{Marayuma1998}
  Y. Yamaguchi and S. Maruyama,
  Chem. Phys. Lett. 286, 336 (1998).

\bibitem{ClassMDFUllForm}
  S. Makino, T. Oda, Y. Hiwatari,
  J.Phys. ChemSolids Vol 58. No. l1.pp. 1845-1851, 1997

\bibitem{ContGrowth}
  Y. Xia, Y. Mu, Y. Xing, R. Wang, C. Tan, L. Mei,
  Phys. Rev. B 57 (1998) 23

\bibitem{Fungimol}
  T. Freeman, \textit{Fungimol, an extensible system for designing atomic-scale objects},
  available at \texttt{http://www.fungible.com/fungimol/index.html}

\end{thebibliography}
\end{document}